\documentclass[11pt]{article}
\usepackage[utf8]{inputenc}
\usepackage[T1]{fontenc}
\usepackage{times}
\usepackage[margin=1in]{geometry}
\usepackage{amsmath,amssymb}
\usepackage{graphicx}
\usepackage{booktabs}
\usepackage{multirow}
\usepackage{array}
\usepackage{longtable}
\usepackage{xcolor}
\usepackage{caption}
\usepackage{subcaption}
\usepackage{url}
\usepackage{xurl}
\usepackage[colorlinks=true,linkcolor=blue!60!black,citecolor=blue!60!black,urlcolor=blue!60!black]{hyperref}
\usepackage[numbers,sort&compress]{natbib}
\usepackage{authblk}
\usepackage{lineno}
\usepackage{setspace}
\usepackage{microtype}
\usepackage{enumitem}
\usepackage{siunitx}
\usepackage{orcidlink}

\graphicspath{{figures/}}
\newcommand{\dataset}[1]{\textsc{#1}}
\newcommand{\ff}{FlightFinder}
\newcommand{\site}{\href{https://himaxym.com}{himaxym.com}}
\newcommand{\apiep}[1]{\texttt{#1}}

\author[1]{Denys Kolomiiets\,\orcidlink{0000-0003-3212-3251}}
\affil[1]{FlightFinder / himaxym.com, Lisbon, Portugal \\ \texttt{denys@himaxym.com}}

\title{Airport-level runway-incursion rates normalized by movements:\\
hierarchical shrinkage estimates, ranking uncertainty and severity\\
for FAA-towered airports, FY2008--FY2024}
\date{\today}

\begin{document}
\maketitle

\begin{abstract}
Runway incursions are one of ICAO's five high-risk occurrence categories, yet airport-level comparisons of incursion risk in the United States still rest on raw counts or on fixed-count thresholds that select on sampling noise. We assemble an airport--fiscal-year panel of 565 FAA-towered airports holding 27{,}079 runway-incursion records from the FAA Runway Safety database as held in the \ff{} corpus (\site), normalized by tower operations from the FAA Terminal Area Forecast (ATADS/OPSNET actuals). A hierarchical negative-binomial model with log operations as an exposure offset, airport random effects, fiscal-year effects and traffic-mix covariates yields, for every airport, a shrinkage-adjusted rate per million operations, an exceedance probability relative to the national median, and a posterior rank with a 90\,\% interval in the manner of Goldstein and Spiegelhalter. Severity (Category A--D) is modeled separately as an ordinal outcome, and empirical-Bayes and fully Bayesian shrinkage are compared. The crude national rate is 29.4 incursions per million operations (95\,\% interval 29.1--29.8), with an index of dispersion of 8.61. Between-airport dispersion after exposure adjustment is $\sigma_u = 1.05$: an airport one standard deviation above the mean carries 2.84 times the national rate. 199 of 565 airports exceed the national median with probability above 0.9, and the 90\,\% rank interval at the median airport spans 134 positions, so much of the league table is statistically indistinguishable from its middle. Severity and frequency respond to different determinants: the airport-level rank correlation between them is $-0.08$ (90\,\% interval $-0.14$ to $-0.02$), so a frequency ranking carries little information about where a severe outcome is likely. The results offer a reproducible, uncertainty-aware alternative to count thresholds for prioritizing runway-safety interventions.
\end{abstract}


\section{Introduction}

A runway incursion is any occurrence at an aerodrome involving the incorrect presence of an aircraft, vehicle or person on the protected area of a surface designated for the landing and take-off of aircraft \citep{faa2026incursions}. Although only a very small fraction of incursions end in a collision, the category retains its place among the highest-consequence event types in commercial aviation because the potential outcome of a single severe event is a high-energy collision between occupied aircraft \citep{eurocontrol2024gappri,stroeve2016strengthening}. In the United States the Federal Aviation Administration (FAA) Runway Safety extract analyzed in this paper records 1{,}679 incursions in fiscal year 2024, nine of them classified in the most severe Categories A and B.\footnote{The FAA's official FY2024 statistic is 1{,}758 incursions \citep{oig2025runway}; the difference reflects the vintage and scope of the Runway Safety file behind our extract (pulled 2026-07-14) relative to the figure the FAA reports, not a filtering choice of this paper: the panel filters of Section~\ref{sec:inclusion} remove only one FY2024 event, leaving 1{,}678.} A recent audit by the Department of Transportation Inspector General found that the FAA lacks an integrated approach to analyzing its own incursion data and that its programs for mitigating airfield geometry deficiencies remain incomplete \citep{oig2025runway}. The national statistics are reported per million operations, but no comparable exposure-normalized measure with uncertainty exists at the level of the individual airport, which is where mitigation decisions, from taxiway reconfiguration to Runway Status Lights, are made.

Existing airport-level evidence is limited in three ways. First, the FAA's Runway Incursion Mitigation (RIM) program identifies problem locations by a raw-count rule: three or more incursions at a location in a calendar year, or an average of one per year over a decade \citep{faa2023rim}. Such thresholds are the canonical case of selection on noise. A location that experiences three events in one year at a low-traffic airport may be entirely unremarkable in expectation, and the subsequent decline that is attributed to mitigation is, in part, regression to the mean \citep{hauer2002estimating,persaud2007empirical}. Second, the academic literature has treated tower operations as one covariate among several rather than as the exposure that converts a count into a rate \citep{mathew2017statistical,johnson2016statistical,omosebi2023investigating}, and it has generally worked with a single cross-section or a short panel, so that between-airport differences cannot be separated from year-to-year sampling variation. Third, severity and frequency have been analyzed in separate studies with different data windows \citep{biernbaum2012runway,wilke2015modelling}, leaving open whether the airports with the highest frequency are also the airports where events are most likely to be severe.

The statistical problem, namely estimating and comparing rates across hundreds of sites when most sites record zero or one event per year, has a mature solution in road safety and in institutional performance comparison. Empirical-Bayes (EB) shrinkage combines an observed count with a model-based prediction so that noisy small-sample rates move toward the group mean by an amount that depends on the estimated overdispersion \citep{hauer2002estimating,efron1975data}; fully Bayesian hierarchical models generalize the same idea and propagate uncertainty in the dispersion and in the exposure itself \citep{lord2010statistical,skaug2026beyond}; and the comparison of sites through rankings must be accompanied by rank intervals because point-estimate league tables built from small counts are dominated by sampling error \citep{goldstein1996league}. None of these tools has previously been applied to the FAA runway-incursion record.

The \ff{} corpus at \site{} holds the FAA Runway Safety incursion set alongside airport identifiers, tower-operation counts and RIM hot-spot locations, within a wider occurrence corpus drawn from 128 source codes, spanning official investigation authorities and compiled databases. Because the incursion set is not exposed through the public API, this paper uses a database export that is byte-identical to the FAA Runway Safety file from which it was loaded; the export and the build scripts are released with the paper so that the panel can be rebuilt from FAA primary sources without access to the site. The airport metadata do use public endpoints.

We ask four questions.
\begin{enumerate}[label=\textbf{RQ\arabic*}]
\item What is the distribution of exposure-normalized runway-incursion rates across FAA-towered airports, and how much of the raw between-airport variation survives adjustment for operations, fiscal-year effects and traffic mix?
\item Which airports have rates that exceed the national median with high posterior probability, and how uncertain are their ranks? How does a shrinkage-based flag compare with the RIM raw-count threshold?
\item Do the determinants of incursion severity (Category A--D) coincide with the determinants of incursion frequency, and do the highest-frequency airports also have the highest probability of a Category A/B outcome?
\item Do airport-geometry covariates, specifically the number of runway crossings and the count of designated hot spots, explain between-airport variation in rates after exposure adjustment, once the endogeneity of hot-spot designation is addressed by lagging?
\end{enumerate}

The contributions are as follows.
\begin{enumerate}
\item A reproducible airport--fiscal-year panel of runway incursions and tower operations for all FAA-towered airports over 17 fiscal years, with documented inclusion rules and a public build script.
\item A hierarchical negative-binomial exposure model that yields, for each airport, a shrinkage-adjusted rate, an exceedance probability and a posterior rank interval, together with a side-by-side comparison of empirical-Bayes and fully Bayesian estimates.
\item An ordinal severity model estimated on the same panel, allowing frequency and severity determinants to be compared directly.
\item An evaluation, specified in advance in the released analysis plan, of the RIM raw-count threshold against the shrinkage-based flag, quantifying how many threshold flags are consistent with sampling noise.
\end{enumerate}

Section~\ref{sec:related} reviews the literature. Section~\ref{sec:data} describes the datasets, the unit of analysis and the exposure denominators. Section~\ref{sec:methods} specifies the frequency, ranking and severity models and the robustness program. Section~\ref{sec:results} is structured by research question, and Sections~\ref{sec:discussion} and~\ref{sec:conclusion} discuss implications and limitations. The appendix contains the reproducibility protocol.

\section{Related work}
\label{sec:related}

\subsection{Definitions, severity classification and official reporting}

The FAA adopts the ICAO definition of a runway incursion (RI) as ``any occurrence at an aerodrome involving the incorrect presence of an aircraft, vehicle or person on the protected area of a surface designated for the landing and take off of aircraft'', and classifies each event into four severity categories, from Category~A (collision narrowly avoided) through Category~D (no immediate safety consequence) \citep{faa2026incursions}. Policy for the collection, classification and review of these events is set by the Runway Safety Program order \citep{faa2013order}, and events are further typed as pilot deviations, operational incidents (controller) or vehicle/pedestrian deviations. The most recent independent audit reports 1{,}758 incursions in FY2024, of which only nine were Category~A/B, and finds that the FAA ``lacks an integrated approach for analyzing runway incursion data'' across its organisational silos \citep{oig2025runway}. The FAA's own Runway Incursion Mitigation programme geo-references incursions to specific airfield locations and flags a location when it records three or more incursions in a calendar year or averages one per year over a decade; by FY2022 it had identified 128 active locations at 82 airports and 6{,}633 non-standard geometry sites \citep{faa2023rim}. Internationally, the European action plan \citep{eurocontrol2017eappri} has been folded into the Global Action Plan for the Prevention of Runway Incursions, a joint EUROCONTROL--ICAO--IATA--ACI--CANSO--FSF document that names runway incursions as one of ICAO's five high-risk occurrence categories \citep{eurocontrol2024gappri}.

\subsection{Statistical analyses of incursion frequency and causes}

Most quantitative RI work uses the FAA Runway Safety database in cross-sectional or short-panel form. \citet{mathew2017statistical} fit count models to US incursions and relate frequency to traffic and airport attributes; \citet{johnson2016statistical} regress incursion counts on the number of runway intersections and taxiways, finding that layout variables explain a substantial share of between-airport variation; and \citet{omosebi2023investigating} update the descriptive picture with post-2015 data, confirming the dominance of pilot deviations and general-aviation involvement. Airport geometry also appears in the Volpe severity study, which analysed roughly 8{,}800 incursions from 2001--2010 with multinomial and ordinal logit models and found that controller-caused events were about three times more likely to be severe, that takeoff-phase events were 2.5 times more likely to be severe than taxi events, and that more runway intersections raised severity whereas more runways lowered it \citep{biernbaum2012runway}. Human-factor studies complement these models: \citet{chang2012human} elicit and rank pilot-, controller- and airport-related risk factors, and \citet{wilke2014airport} propose a holistic taxonomy of surface-operation causal factors from a review of investigation reports. Recent work replaces outcome-based severity with Bayesian-network representations of interacting factors \citep{fu2026interaction}, building on the earlier argument by \citet{stroeve2016strengthening} that RI evaluation should move from observed outcome to modelled collision risk, and on agent-based Monte Carlo collision-risk models of the runway-incursion scenario \citep{stroeve2009systemic}. \citet{wilke2015modelling} model severity directly as an ordinal outcome and show that severity determinants differ from frequency determinants, an important caveat for any rate-based airport comparison.

\subsection{Normalisation by exposure}

Because RI counts scale with traffic, official reporting expresses national totals as incursions per million operations, using FAA tower counts from the Operations Network and its Air Traffic Activity Data System, which record IFR and VFR itinerant and local operations at every federally staffed tower \citep{faa2026opsnet}. Yet almost all airport-level academic analyses either compare raw counts or introduce operations only as a covariate \citep{mathew2017statistical,johnson2016statistical,omosebi2023investigating}, and none, to our knowledge, treats operations as an offset in a model whose primary output is a per-airport rate with uncertainty. The choice of exposure is itself consequential: local (pattern) operations by training aircraft, which are heavily represented among pilot deviations, are counted separately from itinerant operations, and towers at general-aviation reliever airports have different counting practices from Core~30 hubs. Ignoring these differences can move an airport several places in a ranking while leaving the national rate unchanged.

\subsection{Rare-event rates across many sites: empirical Bayes and hierarchical models}

The problem of estimating and ranking rates for many sites with sparse counts has been solved repeatedly outside aviation. In road safety, the empirical Bayes (EB) method combines a site's observed count with a safety-performance-function prediction, shrinking noisy small-sample rates toward a group mean and thereby removing regression-to-the-mean bias from before--after comparisons \citep{hauer2002estimating}. \citet{lord2010statistical} review the count-data families used for this purpose (Poisson, negative binomial, zero-inflated, random-parameter and hierarchical variants) and their pitfalls under low sample mean and overdispersion. The statistical foundation is the Stein shrinkage result of \citet{efron1975data}, and the ranking problem was analysed by \citet{goldstein1996league}, who show that institutional ``league tables'' built from small counts are dominated by sampling error and must be reported with interval estimates and ranking uncertainty. A fully Bayesian hierarchical generalisation of EB that jointly estimates missing exposure and allows the exposure exponent to vary by site class has recently been proposed for crash rates \citep{skaug2026beyond}, and the same structure transfers directly to airports, where operations counts are missing for non-towered fields and the elasticity of incursions to traffic may differ between hub and reliever airports. None of these tools has been applied to the FAA runway-incursion record, although the RIM programme's raw-count thresholds \citep{faa2023rim} are exactly the kind of rule that EB shrinkage was designed to replace.

\subsection{Research gap}

No prior study has produced a multi-year, airport-level panel of runway-incursion rates for FAA-towered airports with at least 5{,}000 operations in the fiscal year in which (i) tower operations from OPSNET/ATADS enter as an explicit exposure offset, (ii) rates are estimated with hierarchical Poisson or negative-binomial shrinkage so that low-volume airports are neither over- nor under-flagged, (iii) severity (A--D) is modelled at the event level on the same panel, with the event type (pilot deviation, operational incident, vehicle/pedestrian deviation) as a covariate, so that separately estimated severity and frequency determinants can be compared directly, and (iv) posterior ranking uncertainty is reported in the manner of \citet{goldstein1996league}. Existing work either models national counts, uses a single cross-section, or relies on raw-count thresholds. The FlightFinder corpus, which holds the FAA runway-incursion set alongside airport identifiers, tower operations and RIM hot-spot locations (the incursion set itself is a database export rather than a public API endpoint), enables this panel to be assembled reproducibly; cross-checking it against the wider occurrence record drawn from 128 source codes (official investigation authorities and compiled sources), so that incursions escalating to reported accidents or serious incidents can be identified at the same airports, is left to future work.

\section{Data}
\label{sec:data}

\subsection{Sources and provenance}

Three datasets are combined (Table~\ref{tab:sources}). The first is the FAA Runway Safety database of runway incursions, which the \ff{} pipeline ingests from the FAA Runway Safety (RWS) export distributed through the ASIAS Runway Incursion query tool and the FAA runway-safety statistics pages \citep{faa2026incursions,faa2013order}. This dataset is not available through the public \site{} API; the analysis uses a database export, \dataset{ri\_faa\_rws}, whose row count, checksum and FAA file date are recorded in the appendix (35{,}308 records spanning FY2002--FY2026, of which 27{,}242 fall in the analysis window, extracted 2026-07-14, SHA-256 \texttt{0cba71db\dots}). Each record carries an event identifier, the incursion date, the airport identifier, the event type (pilot deviation, operational incident or vehicle/pedestrian deviation), the severity category (A--D, with a separate code for events not yet categorized or categorized as ``E'' in some FAA releases), the runway or intersection involved, the aircraft or vehicle types and a short narrative. The FAA has applied the ICAO definition and the four-category severity scale since FY2008 \citep{faa2013order}; the FAA's own statistics distinguish pre- and post-FY2008 series, and so do we (Section~\ref{sec:inclusion}).

The second dataset is airport operations from the FAA Terminal Area Forecast (TAF) bulk release (\texttt{APO100\_TAF\_Final\_2025.zip}), whose historical years are the ATADS/OPSNET tower counts \citep{faa2026opsnet}, published per airport and fiscal year with the split the model needs: itinerant operations by user class (air carrier \texttt{itn\_Ac}, air taxi \texttt{itn\_at}, general aviation \texttt{itn\_ga}, military \texttt{itn\_mil}) and local operations (\texttt{loc\_ga}, \texttt{loc\_mil}). Two properties of the TAF tables matter for reproduction: each airport--year appears under several scenario codes, and only the \texttt{scenario}~$=0$ rows are the actual/baseline series (the others are forecast variants that would multiply-count exposure); and the location identifiers carry trailing whitespace that must be trimmed before joining. We use airport operations, the sum of itinerant and local operations, as the primary exposure measure, and itinerant operations alone as an alternative. The third dataset is airport metadata: the towered-airport flag (\texttt{ATCT\_FLAG}) and hub class (\texttt{HUB\_SIZE}) from the TAF airport inventory, supplemented by the public endpoint \apiep{GET /airports/\{ident\}} for identifiers and coordinates. Geometry covariates (runway crossings, hot spots) were planned from the FAA Airport Data and Information Portal and the RIM annual inventory \citep{faa2023rim} but were not obtained; RQ4 is therefore not estimated (Section~\ref{sec:results}), and only the RIM location list, extracted from the July 2026 inventory PDF, is used, for the flag comparison of RQ2.

\begin{table}[htbp]
\centering
\caption{Data sources.}
\label{tab:sources}
\footnotesize
\begin{tabular}{@{}p{2.5cm}p{4.2cm}p{2.3cm}p{2.1cm}p{1.9cm}p{1.5cm}@{}}
\toprule
Dataset & Provenance & Access & Window & Unit & Rows \\
\midrule
\dataset{ri\_faa\_rws} & FAA Runway Safety / RWS export \citep{faa2026incursions} & DB export (not in API) & FY2008--FY2024 & incursion event & 27{,}242 \\
\dataset{ops\_taf} & FAA Terminal Area Forecast actuals (ATADS/OPSNET basis) \citep{faa2026opsnet} & FAA bulk download & FY2008--FY2024 & airport--fiscal-year & 9{,}582 \\
Airport metadata & FAA Terminal Area Forecast inventory & FAA bulk download & current & airport & 565 \\
RIM location inventory & FAA portal; PDF table \citep{faa2023rim} & PDF extraction & validated 2015--2025 & airport location & 134 (77 airports) \\
\dataset{adip\_geometry} & FAA ADIP runway crossings & \emph{not obtained} & --- & airport & --- \\
\bottomrule
\end{tabular}
\end{table}

\subsection{Unit of analysis and coverage window}

The unit of analysis is the airport--fiscal-year (FAA fiscal years run from 1 October to 30 September, matching the FAA's own reporting). The panel comprises every airport carrying an FAA or federal-contract tower in the Terminal Area Forecast inventory (\texttt{ATCT\_FLAG} set) and recording at least 5{,}000 operations in the fiscal year, giving 565 airports and 9{,}582 airport--years over FY2008--FY2024. Exposure is published annually rather than monthly in this release, so the ``at least ten reported months'' rule of the analysis plan is replaced by the minimum-operations threshold, which is varied from 1{,}000 to 20{,}000 in the sensitivity analysis. Incursions recorded at airports with no tower exposure record, or below the threshold, have no observed denominator: 163 events, 0.6\,\% of those in the window, are excluded from the primary panel; a latent-exposure treatment in the spirit of \citet{skaug2026beyond} is specified in the analysis plan but was not run in this version (Section~\ref{sec:robust-results}). The window closes at FY2024 because Terminal Area Forecast actuals end there; the 1{,}477 incursions recorded in FY2025 are held out rather than divided by forecast operations.

\subsection{Variables}

Table~\ref{tab:variables} lists the analysis variables. The outcome for the frequency model is $y_{it}$, the number of incursions at airport $i$ in fiscal year $t$, with type-specific counts $y_{it}^{\mathrm{PD}}$, $y_{it}^{\mathrm{OI}}$ and $y_{it}^{\mathrm{VPD}}$ and severity-specific counts $y_{it}^{\mathrm{AB}}$ and $y_{it}^{\mathrm{CD}}$. The outcome for the severity model is the ordinal category of each event. The exposure $E_{it}$ is airport operations in the fiscal year. The estimated airport-level covariates are the general-aviation share of airport operations (itinerant plus local GA over total operations), the local (pattern) share of airport operations, and the hub class. Further covariates named in the analysis plan --- tower type, surface-surveillance coverage (ASDE-X or ASSC) and Runway Status Lights indicators \citep{oig2025runway}, the number of runways, the number of runway crossings, and the lagged count of RIM hot spots --- were not estimated, because the required extracts were not obtained (Section~\ref{sec:limitations}).

\begin{table}[htbp]
\centering
\caption{Variables. Source fields for \dataset{ri\_faa\_rws} are the column names of the frozen export header; exposure fields are those of the TAF \texttt{AirportsOperations} table (\texttt{scenario} $=0$ rows). The column crosswalk is released as \texttt{schema/ri\_columns.yaml}.}
\label{tab:variables}
\footnotesize
\begin{tabular}{@{}lllll@{}}
\toprule
Variable & Symbol & Source field & Level & Role \\
\midrule
Event identifier & -- & \texttt{EVENT\_ID} & event & key, dedup \\
Incursion count & $y_{it}$ & events per (\texttt{EVENT\_ARPT\_ID}, FY) & airport--FY & outcome \\
Event type & -- & \texttt{INCDNT\_TYPE\_FAA\_CODE} & event & stratifier \\
Severity category & $S_k$ & \texttt{RWY\_SFTY\_RI\_CAT\_RNK\_CODE} (A/B/C/D) & event & ordinal outcome \\
Operator of first aircraft & $\mathbf{z}_k$ & \texttt{ACFT\_1\_FLTCNDT\_CODE} & event & severity covariate \\
Airport identifier & $i$ & \texttt{EVENT\_ARPT\_ID}; \texttt{LOCID} (TAF) & event/airport & key \\
Fiscal year & $t$ & derived from \texttt{EVENT\_LCL\_DATE} & airport--FY & key \\
Airport operations & $E_{it}$ & sum of the \texttt{itn\_*} and \texttt{loc\_*} fields & airport--FY & offset \\
Itinerant operations & $E^{\mathrm{itin}}_{it}$ & sum of the four \texttt{itn\_*} fields & airport--FY & alt.\ offset \\
GA share & $x_{1,it}$ & (\texttt{itn\_ga}+\texttt{loc\_ga})/$E_{it}$ & airport--FY & covariate \\
Local share & $x_{2,it}$ & (\texttt{loc\_ga}+\texttt{loc\_mil})/$E_{it}$ & airport--FY & covariate \\
Hub class & $x_{3,i}$ & \texttt{HUB\_SIZE} (TAF) & airport & covariate \\
Towered-airport flag & -- & \texttt{ATCT\_FLAG} (TAF) & airport & inclusion \\
RIM airport flag & -- & RIM inventory \citep{faa2023rim} & airport & RQ2 comparison \\
\bottomrule
\end{tabular}
\end{table}

\subsection{Inclusion and exclusion criteria}
\label{sec:inclusion}

The primary analysis window begins in FY2008, when the FAA adopted the ICAO definition and reclassified many former surface incidents into Categories C and D \citep{faa2013order}; the pre-FY2008 series is used only in a change-point robustness check. Events enter the frequency count if they carry a valid airport identifier that matches a TAF-reporting towered airport--year in the panel, \emph{regardless} of type or severity code: the outcome $y_{it}$ counts all in-window events, including the 3.9\,\% with legacy or uncoded type codes (\texttt{OD}, \texttt{OTH}, \texttt{UNK}) and the 13.1\,\% with no severity category. The severity model retains only events with a category in \{A, B, C, D\} \emph{and} a coded PD, OI (pooling the legacy \texttt{OE} code) or VPD type; the type-specific frequency models count PD, OI(+\texttt{OE}) and VPD events. A sensitivity analysis that restricts the frequency outcome to the coded PD/OI/VPD types, which matters because the uncoded types concentrate in FY2016--FY2017, is reported in Section~\ref{sec:robust-results}. Duplicate records, identified by identical \texttt{EVENT\_ID} or by identical (airport, date, type, runway) tuples, are collapsed with the rule in the appendix. Airport--years with fewer than 5{,}000 operations are excluded because the offset becomes unstable; the threshold is varied in a sensitivity analysis.

\subsection{Known biases}

Four biases are anticipated. Reporting and classification drift: the FY2008 definitional change, subsequent revisions to Order 7050.1 and the FY2023 runway-safety push following several high-profile events plausibly changed the propensity to record and to categorize C/D events; fiscal-year effects absorb national shifts but not airport-specific ones. Detection: surface surveillance both reduces risk and increases the probability that a minor incursion is detected, and coverage correlates with hub size, so the surveillance coefficient is not a treatment effect. Exposure mismeasurement: contract towers and FAA towers count operations under slightly different practices, local operations are estimated at some facilities, and the elasticity of incursions to operations need not be unity; the exposure exponent is therefore estimated as well as fixed. Traffic mix: pilot deviations are concentrated among general-aviation and training operations, so a rate that ignores GA share partly measures the presence of flight schools.

\section{Methods}
\label{sec:methods}

\subsection{Frequency model with exposure offset}

Let $y_{it}$ be the incursion count at airport $i=1,\dots,I$ in fiscal year $t=1,\dots,T$ and $E_{it}$ the airport operations. The primary model is a hierarchical negative-binomial (NB2) regression,
\begin{equation}
y_{it} \sim \mathrm{NegBin}(\mu_{it}, \phi), \qquad
\mathbb{E}[y_{it}] = \mu_{it}, \qquad
\mathrm{Var}[y_{it}] = \mu_{it} + \mu_{it}^2/\phi,
\label{eq:nb}
\end{equation}
\begin{equation}
\log \mu_{it} = \log E_{it} + \alpha + u_i + \gamma_t + \mathbf{x}_{it}^{\top}\boldsymbol{\beta},
\qquad u_i \sim \mathcal{N}(0, \sigma_u^2),
\label{eq:linpred}
\end{equation}
where $\log E_{it}$ enters as an offset with coefficient fixed at one, $u_i$ is an airport random effect, $\gamma_t$ are fiscal-year random intercepts, $\gamma_t \sim \mathcal{N}(0, \sigma_\gamma^2)$, and $\mathbf{x}_{it}$ collects the covariates in Table~\ref{tab:variables}. The airport-specific rate per operation is $\lambda_{it} = \mu_{it}/E_{it} = \exp(\alpha + u_i + \gamma_t + \mathbf{x}_{it}^{\top}\boldsymbol{\beta})$, and the quantity reported per airport is the time-averaged rate $\lambda_i = \exp(\alpha + u_i + \bar{\mathbf{x}}_i^{\top}\boldsymbol{\beta})$ scaled to incursions per million operations. The NB2 form is the standard choice for overdispersed count panels \citep{lord2010statistical} and nests the Poisson as $\phi \to \infty$. Following \citet{skaug2026beyond} we also estimate a version in which the exposure exponent is free and varies by hub class,
\begin{equation}
\log \mu_{it} = \rho_{c(i)} \log E_{it} + \alpha + u_i + \gamma_t + \mathbf{x}_{it}^{\top}\boldsymbol{\beta},
\label{eq:elastic}
\end{equation}
and test $\rho_c = 1$ for each class $c$; a rejection indicates that incursions scale sub- or super-linearly with traffic and that the offset model misranks airports at the extremes of the traffic distribution. Priors are weakly informative: $\alpha \sim \mathcal{N}(-10.4, 1.5^2)$, centered on the published national rate ($\log(10^{-6}\cdot 29.4) \approx -10.4$), $\boldsymbol{\beta} \sim \mathcal{N}(0, 1)$, $\sigma_u \sim \mathrm{half}\text{-}\mathcal{N}(0,1)$, $\phi \sim \mathrm{Gamma}(2, 0.5)$ and $\gamma_t \sim \mathcal{N}(0, \sigma_\gamma^2)$ with $\sigma_\gamma \sim \mathrm{half}\text{-}\mathcal{N}(0,1)$. The intercept prior is set explicitly in the primary, type-specific and robustness-refit models; in the free-exponent model of Eq.~\eqref{eq:elastic} and its coded-types variant, the zero-inflated and hurdle families, and the extended-panel fit, the intercept keeps the \texttt{brms} default (a weakly informative Student-$t$ prior centered near the data), a deviation from a uniform prior specification that we note here rather than leave implicit. Sensitivity to the prior scale is examined in Section~\ref{sec:robust}.

The same specification is estimated separately for the three event types. Category A/B counts are too sparse at the airport level to support a count model or a ranking (nine events nationwide in FY2024; \citealt{oig2025runway}); severity enters through the event-level models of Section~\ref{sec:methods}, including a binary A/B-versus-C/D logistic specification reported at the level of covariate effects only. The analysis plan also names an airport-level model of Category C/D counts (its M7); that model was not estimated in this version, and C/D events enter only through the all-events count and the event-level severity models.

\subsection{Empirical-Bayes shrinkage and comparison with the full posterior}

The empirical-Bayes estimator of \citet{hauer2002estimating} is computed as a benchmark. With $\hat{\mu}_i = \sum_t \hat{\mu}_{it}^{\mathrm{SPF}}$ the count predicted for airport $i$ over the panel from the fixed-effects part of Eq.~\eqref{eq:linpred} (the safety-performance function) and $y_i = \sum_t y_{it}$ the observed total, the EB expected count is
\begin{equation}
\hat{m}_i^{\mathrm{EB}} = w_i \hat{\mu}_i + (1-w_i) y_i, \qquad w_i = \frac{1}{1 + \hat{\mu}_i/\hat{\phi}},
\label{eq:eb}
\end{equation}
with variance $\mathrm{Var}[\hat{m}_i^{\mathrm{EB}}] = (1-w_i)\hat{m}_i^{\mathrm{EB}}$, and the EB rate is $\hat{m}_i^{\mathrm{EB}} / \sum_t E_{it}$. The weight $w_i$ falls toward zero for high-traffic airports, whose observed counts are informative, and rises toward one for low-traffic airports, whose counts are dominated by noise. The EB estimate is compared with the posterior mean of $\lambda_i$ from the full hierarchical model; the two coincide asymptotically \citep{efron1975data} but differ when $\phi$ and $\sigma_u$ are uncertain, when covariates are few and when exposure is itself estimated \citep{skaug2026beyond}. The comparison is reported as a scatter of EB against fully Bayesian rates with the number of airports whose exceedance flag changes between methods.

\subsection{Exceedance probabilities and posterior ranks}

For each airport we report the posterior exceedance probability
\begin{equation}
\pi_i = \Pr\big(\lambda_i > \lambda_{\mathrm{med}} \,\big|\, \mathbf{y}\big), \qquad \lambda_{\mathrm{med}} = \mathrm{median}_j\, \lambda_j,
\label{eq:exceed}
\end{equation}
where the median is recomputed within each posterior draw, and the posterior distribution of the rank
\begin{equation}
R_i = 1 + \sum_{j \neq i} \mathbb{I}(\lambda_j < \lambda_i),
\label{eq:rank}
\end{equation}
so that ranks ascend with the rate --- the airport with the highest rate among $I$ airports has rank $I$ --- summarized by its expectation and a 90\,\% equal-tailed interval \citep{goldstein1996league}. A shrinkage-based flag is defined as $\pi_i \geq 0.90$; the comparison with the RIM raw-count rule, specified in advance in the analysis plan, counts airports (RIM operates at the level of airfield locations; airport-level RIM status is defined here as having at least one location in the RIM inventory, mapped to the airport by the prefix of the RIM location identifier --- all 77 RIM airports matched a panel airport) that are flagged by one rule and not the other, and reports the fraction of RIM-flagged airports whose 90\,\% rank interval includes the median rank. Ranks are reported on the national scale only; a within-hub-class ranking, though natural given that a national ranking mixes airports whose exposure differs by two orders of magnitude, is not computed in this version.

\subsection{Severity model}

Each event $k$ at airport $i(k)$ in fiscal year $t(k)$ has an ordinal severity $S_k \in \{\mathrm{D} < \mathrm{C} < \mathrm{B} < \mathrm{A}\}$. We fit a cumulative-logit (proportional-odds) model with an airport random intercept,
\begin{equation}
\Pr(S_k \leq s) = \mathrm{logit}^{-1}\big(\tau_s - \eta_k\big), \qquad
\eta_k = v_{i(k)} + \gamma^{S}_{t(k)} + \mathbf{z}_k^{\top}\boldsymbol{\zeta}, \qquad v_i \sim \mathcal{N}(0, \sigma_v^2),
\label{eq:ordinal}
\end{equation}
with thresholds $\tau_{\mathrm{D}} < \tau_{\mathrm{C}} < \tau_{\mathrm{B}}$ and event-level covariates $\mathbf{z}_k$ comprising the event type (PD, OI with the legacy \texttt{OE} code pooled, VPD) and the operating rule of the first aircraft (part 91/121/129/135, vehicle, pedestrian, unknown) --- the two predictors available in the source export. Phase of flight and the aircraft-class pair, named in the analysis plan following the variable sets of \citet{biernbaum2012runway} and \citet{wilke2015modelling}, are not recorded in the export and were not estimated (Section~\ref{sec:limitations}); no airport-level covariates enter $\mathbf{z}_k$, and airport and fiscal-year heterogeneity is carried by the random intercepts $v_i$ and $\gamma^{S}_{t}$ alone. Because Categories A and B are rare, the proportional-odds specification is checked against a binary A/B-versus-C/D logistic model; the partial-proportional-odds alternative named in the analysis plan is not run in this version. RQ3 is answered by the draw-by-draw posterior rank correlation between the frequency random effects $u_i$ and the severity random effects $v_i$ from the two separately estimated models (the bivariate joint random-effects model named in the plan as a secondary specification was not run), and by comparing the signs and magnitudes of $\boldsymbol{\beta}$ and $\boldsymbol{\zeta}$ for shared covariates.

\subsection{Airport-geometry covariates}

Runway crossings and hot-spot counts enter the frequency model as a separate block so that their incremental explanatory power can be reported as the reduction in $\sigma_u^2$ and the change in expected log predictive density. Hot-spot designation is a response to past incursions \citep{faa2023rim,johnson2016statistical}; to avoid reverse causality the hot-spot count is lagged by one fiscal year and, in a robustness check, fixed at its value in the first RIM inventory year available. Runway crossings are time-invariant over the panel except where a documented reconfiguration occurred. This covariate block was not estimated in this version: the ADIP runway-crossing extract and the year-by-year RIM hot-spot histories were not obtained, and Section~\ref{sec:results} states the consequence for RQ4 rather than substituting a proxy.

\subsection{Estimation, inference and model checking}

Models are estimated by Hamiltonian Monte Carlo in Stan \citep{carpenter2017stan} through the \texttt{brms} interface; the primary model uses four chains of 2{,}000 sampling iterations after 1{,}000 warm-up, and the chain settings of every other fit are listed in the appendix (Step~8). Convergence diagnostics: all parameters of the frequency models reach $\hat{R} \leq 1.01$ (two hub-class coefficients in the primary fit attain 1.01 exactly), while in the severity model the first cumulative threshold reaches $\hat{R} = 1.02$ with a bulk effective sample size of 547 --- adequate for the point estimate but below the 100-per-chain heuristic that triggers Stan's low-ESS warning, so the severity intervals are slightly less precise than the nominal draw count suggests. Bulk and tail effective sample sizes are above 400 throughout, and no fit produced divergent transitions after reparameterization. Model comparison uses leave-one-out cross-validation with Pareto-smoothed importance sampling, and adequacy is checked with posterior predictive distributions of the proportion of zero airport--years, the maximum count and the dispersion index. For the EB benchmark the safety-performance function is fitted by maximum likelihood. All intervals are 90\,\% posterior intervals unless stated otherwise; no null-hypothesis tests are reported for the ranking quantities, which are inherently descriptive.

\subsection{Robustness and sensitivity analyses specified in advance}
\label{sec:robust}

The following analyses are fixed in advance in the analysis plan released with the paper; the plan is a repository document, not a registry entry. (i) Distribution family: Poisson, NB2, zero-inflated NB and hurdle NB are compared by LOO, and the NB2 is retained unless the zero-inflated model improves expected log predictive density by more than 4 units with a standard error smaller than the difference. (ii) Exposure definition: airport operations versus itinerant operations versus type-specific GA exposure. (iii) Exposure exponent: Eq.~\eqref{eq:elastic} versus the offset model. (iv) Window: exclusion of FY2020--FY2021, when air-carrier operations collapsed far more than general-aviation operations, and exclusion of the most recent fiscal year, in which severity classification may be incomplete. (v) Change point: extension of the panel backward from FY2008 with a step in the intercept at FY2008 (the plan names FY2000 as the lower bound; the frozen extract begins in October 2001, so the extension runs from FY2002 --- fiscal years FY2000--FY2001 have no incursion records at all and would enter the model as fabricated zeros). (vi) Minimum-operations threshold for airport--year inclusion. (vii) Prior scale doubled and halved for $\sigma_u$ and $\phi$. (viii) Leave-one-airport-out influence on the national rate and on $\sigma_u$. (ix) Latent exposure for non-towered airports following \citet{skaug2026beyond}, with operations imputed from based-aircraft counts. Any change in the set of airports with $\pi_i \geq 0.90$ under these variations is reported as a stability table.

\subsection{Software and data handling}

The frozen extract, the exposure panel and the severity dataset are built with Node.js 26.7 scripts released alongside the paper (\texttt{build\_descriptives.mjs}, \texttt{build\_panel.mjs}, \texttt{build\_panel\_2000.mjs}, \texttt{build\_severity.mjs}, \texttt{build\_rim.mjs}); they are deterministic and take no sample. Models are fitted in R 4.6.1 with \texttt{brms} 2.23.0 on the \texttt{rstan} 2.32.7 backend, and compared with \texttt{loo} 2.10.1. The random seed is 20260827; package versions and the checksums of the frozen extracts are listed in the appendix.

\section{Results}
\label{sec:results}

\subsection{Descriptive overview}

The frozen extract contains 27{,}242 incursions at 525 distinct location identifiers over the 17 fiscal years FY2008--FY2024; 515 of those identifiers match an airport in the towered exposure panel, and the ten that do not (an empty identifier, three air-traffic facilities --- \texttt{HCF}, \texttt{I90}, \texttt{ZKC} --- and six airfield identifiers with no qualifying TAF exposure record) carry 163 events between them. Counted with the same type definitions as Table~\ref{tab:descriptives}, 59.7\,\% of the extract's events are pilot deviations, 16.3\,\% operational incidents (pooling the legacy \texttt{OE} code) and 20.1\,\% vehicle/pedestrian deviations; the remaining 3.9\,\% carry legacy or uncoded type codes discussed below. Of all events, 34.9\,\% are Category C, 50.9\,\% Category D and 239 events (0.9\,\%) are Category A or B; 13.1\,\% are not categorized. Joined to exposure, the panel covers 9{,}582 airport--years at 565 towered airports and retains 27{,}079 incursions, 99.4\,\% of those in the window; the 163 events that fall outside are those at the ten unmatched identifiers above. Across the panel 40.8\,\% of airport--years record zero incursions, the sample mean count is 2.83 and the sample variance is 24.34, an index of dispersion of 8.61 --- far above the Poisson benchmark and above the threshold of 1.5 specified in advance in the released analysis plan, so the negative-binomial specification is retained as primary. Total airport operations in the panel are 919.7 million, giving a crude national rate of 29.4 incursions per million operations (95\,\% Poisson interval 29.1--29.8; Table~\ref{tab:descriptives}). The annual rate is not flat: it rises from 18.8 per million in FY2008 to a peak of 47.5 in FY2017, falls back to 26.6 in the pandemic year FY2020 and settles near 29 thereafter. The FY2016--FY2017 spike deserves scrutiny before it is read as a change in occurrence: 311 and 333 of the panel events in those two fiscal years carry type codes outside PD/OI/VPD (almost entirely \texttt{OTH} and \texttt{UNK}), against at most 53 in any other year, so part of the spike is a coding artefact rather than a documented reporting expansion. The sensitivity analysis of Section~\ref{sec:robust-results} therefore refits the frequency model with the outcome restricted to the coded types.

The type coding is not stable across the window. Operational incidents appear under the code \texttt{OI} only from FY2012; in FY2008--FY2011 the same events are coded \texttt{OE} (operational error), the FAA's pre-2012 terminology. Pooling the two codes is therefore required for any series that spans the change, and the apparent FY2012 jump in operational incidents in unpooled counts is an artefact of the recoding rather than a change in occurrence.

\begin{table}[htbp]
\centering
\caption{Panel descriptives by fiscal year. Airports are those in the exposure panel (towered, at least 5{,}000 annual operations); the total row gives distinct airports over the window, not a column sum. All counts are panel counts, i.e.\ events joined to an exposure record; PD, OI and VPD do not sum to the Incursions column because 1{,}062 panel events (3.9\,\%) carry legacy or uncoded type codes (Section~\ref{sec:inclusion}). OI pools the \texttt{OI} and legacy \texttt{OE} codes. Operations are FAA Terminal Area Forecast actuals; Rate is incursions per million operations.}
\label{tab:descriptives}
\small
\begin{tabular}{@{}lrrrrrrrr@{}}
\toprule
FY & Airports & Incursions & PD & OI & VPD & A/B & Ops (M) & Rate \\
\midrule
2008 & 562 & 1{,}181 & 723 & 140 & 292 & 24 & 62.8 & 18.8 \\
2009 & 562 & 1{,}124 & 690 & 115 & 277 & 11 & 56.6 & 19.8 \\
2010 & 563 & 1{,}180 & 708 & 130 & 309 & 6 & 54.6 & 21.6 \\
2011 & 561 & 1{,}290 & 682 & 308 & 266 & 7 & 54.0 & 23.9 \\
2012 & 564 & 1{,}351 & 813 & 217 & 295 & 18 & 53.3 & 25.3 \\
2013 & 564 & 1{,}512 & 893 & 256 & 332 & 11 & 52.6 & 28.8 \\
2014 & 564 & 1{,}621 & 935 & 267 & 391 & 14 & 52.0 & 31.2 \\
2015 & 564 & 1{,}754 & 1{,}024 & 339 & 384 & 15 & 52.2 & 33.6 \\
2016 & 564 & 2{,}184 & 1{,}129 & 336 & 408 & 18 & 52.3 & 41.7 \\
2017 & 564 & 2{,}506 & 1{,}367 & 341 & 465 & 8 & 52.7 & 47.5 \\
2018 & 564 & 1{,}794 & 1{,}115 & 340 & 329 & 12 & 54.2 & 33.1 \\
2019 & 564 & 1{,}741 & 1{,}110 & 321 & 294 & 12 & 55.7 & 31.3 \\
2020 & 564 & 1{,}244 & 825 & 164 & 240 & 15 & 46.8 & 26.6 \\
2021 & 564 & 1{,}560 & 1{,}024 & 226 & 280 & 18 & 49.9 & 31.2 \\
2022 & 564 & 1{,}730 & 1{,}083 & 309 & 312 & 16 & 54.7 & 31.6 \\
2023 & 565 & 1{,}629 & 1{,}009 & 313 & 266 & 22 & 56.5 & 28.8 \\
2024 & 565 & 1{,}678 & 1{,}032 & 294 & 299 & 9 & 58.7 & 28.6 \\
\midrule
Total & 565 & 27{,}079 & 16{,}162 & 4{,}416 & 5{,}439 & 236 & 919.7 & 29.4 \\
\bottomrule
\end{tabular}
\end{table}

\begin{figure}[htbp]
\centering
\includegraphics[width=0.49\linewidth]{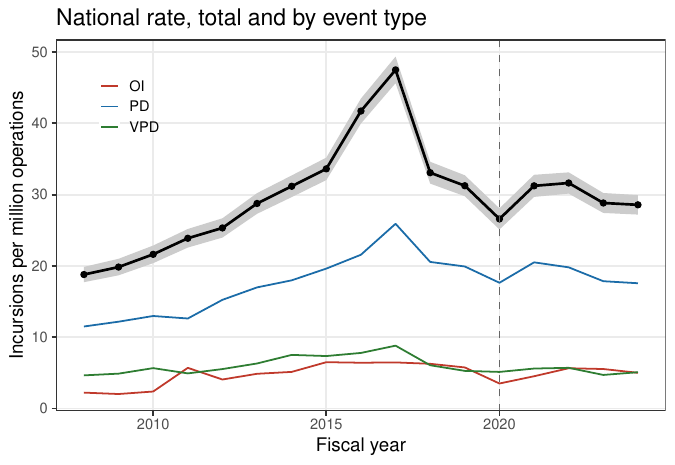}\hfill
\includegraphics[width=0.49\linewidth]{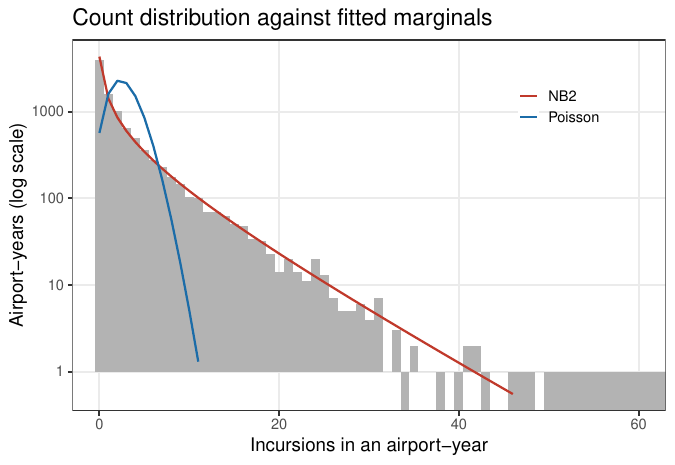}
\caption{National runway-incursion rate by fiscal year and event type (left), and the distribution of airport--year counts with fitted Poisson and negative-binomial marginals (right). The excess of zeros and the long right tail relative to the Poisson motivate the negative-binomial specification.}
\label{fig:national}
\end{figure}

\subsection{RQ1: distribution of exposure-normalized rates}

The hierarchical NB2 model estimates a national mean rate of 14.30 per million operations (90\,\% interval 11.38--18.17) for a non-hub airport at mean traffic mix, a dispersion parameter $\phi = 2.99$ and a between-airport standard deviation $\sigma_u = 1.05$ (0.98--1.11) on the log scale, implying that an airport one standard deviation above the mean has a rate 2.84 times the national mean after adjustment. The raw between-airport coefficient of variation of crude rates is 0.94; after shrinkage it is 0.89. Covariate effects are given in Table~\ref{tab:frequency}. As a rate ratio the general-aviation share carries 3.58 (2.77--4.65) per unit of share, the local share 0.11 (0.08--0.15), and the hub-class contrasts against non-hub airports are 1.56 (1.24--1.94) for small, 1.57 (1.13--2.13) for medium and 1.57 (1.09--2.26) for large hubs. The free exposure exponent $\rho_c$ is 0.80 (0.66--0.94) for large hubs, 0.83 (0.68--0.98) for medium hubs, 0.73 (0.59--0.87) for small hubs and 0.68 (0.59--0.77) for non-hub towered airports. None of the four intervals contains 1, so the offset assumption is rejected in every hub class: incursions scale \emph{sub-linearly} with traffic, and a rate expressed per million operations therefore flatters busy airports and penalizes quiet ones. This matters for how the results are used. Under a fixed offset the ranking is a ranking of rates; under $\rho < 1$ part of what separates airports at the extremes of the traffic distribution is the exposure specification rather than safety performance. We report the offset model as primary because it is the specification fixed in advance in the released analysis plan and the one whose units are comparable to the FAA's published statistics, but the free-exponent estimates are the reason the airport rankings in Section~\ref{sec:results} should not be read as a like-for-like comparison between a large hub and a non-towered field.

Posterior predictive checks support the negative binomial on the two quantities the sparse-count literature warns about and expose one limit. The model reproduces the share of zero airport--years (38.6\,\% replicated against 40.8\,\% observed) and the index of dispersion (9.73 against 8.61), but it under-predicts the maximum count: the largest observed airport--year holds 171 incursions while the replicated maximum averages 90. The upper tail is therefore heavier than a negative binomial with a common shape parameter can generate, which argues for a shape that varies by hub class in future work and means the very highest-count airports are fitted less well than the body of the distribution. The same caveat applies to the LOO comparison with the Poisson in Table~\ref{tab:frequency}: the Pareto-smoothed importance sampling behind it flags 9 observations with $\hat{k} > 0.7$ under the NB2 and 16 under the Poisson, so the $\Delta$ELPD of $+1509.5$ is an approximate magnitude rather than an exact one --- the direction of the comparison is not in doubt given its size relative to the standard error, but no moment matching or $K$-fold refit was run.

\begin{table}[htbp]
\centering
\caption{Hierarchical negative-binomial frequency models: posterior medians and 90\,\% intervals of rate ratios (exponentiated coefficients). Column (1) is the primary offset model, (2) the free-exponent model of Eq.~\eqref{eq:elastic}, (3)--(5) type-specific models. The LOO comparison with the Poisson is PSIS-based and approximate: 9 observations exceed Pareto $\hat{k} = 0.7$ under the NB2 and 16 under the Poisson.}
\label{tab:frequency}
\footnotesize
\setlength{\tabcolsep}{3pt}
\begin{tabular}{@{}lccccc@{}}
\toprule
& (1) All & (2) All, free $\rho$ & (3) PD & (4) OI & (5) VPD \\
\midrule
Intercept (per $10^6$ ops) & 14.30 (11.38--18.17) & -- & 6.45 (4.98--8.33) & 1.68 (1.18--2.33) & 4.60 (3.53--5.93) \\
Exposure exponent $\rho$ (large hub) & 1 & 0.80 (0.66--0.94) & 1 & 1 & 1 \\
Exposure exponent $\rho$ (medium hub) & 1 & 0.83 (0.68--0.98) & 1 & 1 & 1 \\
Exposure exponent $\rho$ (small hub) & 1 & 0.73 (0.59--0.87) & 1 & 1 & 1 \\
Exposure exponent $\rho$ (non-hub) & 1 & 0.68 (0.59--0.77) & 1 & 1 & 1 \\
GA share (rate ratio) & 3.58 (2.77--4.65) & -- & 4.30 (3.19--5.81) & 1.62 (1.10--2.44) & 2.80 (1.98--3.94) \\
Local share (rate ratio) & 0.11 (0.08--0.15) & -- & 0.13 (0.09--0.19) & 0.21 (0.13--0.36) & 0.12 (0.08--0.18) \\
Small hub (vs non-hub) & 1.56 (1.24--1.94) & -- & 1.82 (1.42--2.33) & 2.19 (1.71--2.84) & 0.92 (0.74--1.16) \\
Medium hub (vs non-hub) & 1.57 (1.13--2.13) & -- & 1.72 (1.18--2.47) & 2.87 (2.00--4.17) & 1.04 (0.75--1.43) \\
Large hub (vs non-hub) & 1.57 (1.09--2.26) & -- & 1.48 (0.98--2.26) & 4.38 (2.95--6.63) & 0.88 (0.61--1.25) \\
\multicolumn{6}{@{}l@{}}{\footnotesize Surveillance and geometry covariates await the ADIP and RIM extracts; see Section~\ref{sec:limitations}.} \\
\midrule
$\sigma_u$ & 1.05 (0.98--1.11) & -- & 1.14 (1.06--1.22) & 0.98 (0.89--1.07) & 0.91 (0.85--0.98) \\
$\phi$ & 2.99 (2.81--3.20) & -- & 2.53 (2.35--2.74) & 1.84 (1.62--2.08) & 1.47 (1.33--1.64) \\
$\sigma_\gamma$ & 0.27 (0.20--0.38) & -- & 0.24 (0.18--0.34) & 0.39 (0.29--0.55) & 0.18 (0.13--0.27) \\
ELPD (LOO) vs Poisson & $+1509.5$ (SE 250.5) & -- & -- & -- & -- \\
Airport--years & 9{,}582 & 9{,}582 & 9{,}582 & 9{,}582 & 9{,}582 \\
\bottomrule
\end{tabular}
\end{table}

\begin{figure}[htbp]
\centering
\includegraphics[width=\linewidth]{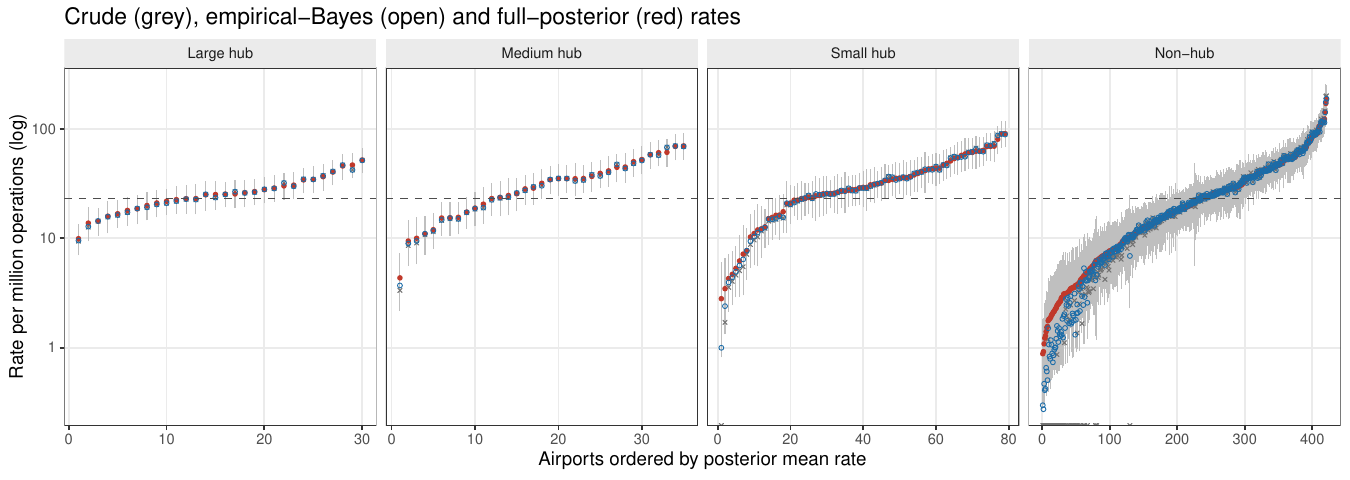}
\caption{Crude, empirical-Bayes and fully Bayesian airport-level incursion rates per million operations, by hub class. Shrinkage toward the class mean is strongest for low-volume towers, where the crude rate is dominated by one or two events; intervals are 90\,\% posterior intervals.}
\label{fig:caterpillar}
\end{figure}

\subsection{RQ2: exceedance and rank uncertainty; comparison with the RIM threshold}

199 airports have $\pi_i \geq 0.90$ and 209 have $\pi_i \leq 0.10$; the remaining 157 cannot be distinguished from the median at this level. The width of the 90\,\% rank interval is 134 positions at the median airport, out of 565 ranked airports (Table~\ref{tab:ranks}). A quarter of the league table is therefore indistinguishable from the median airport's position, which is the central caution against reading any point rank as a finding.

The composition of the top of the table is itself a result, and a cautionary one: all fifteen airports with the highest exceedance probability are non-hub fields (Table~\ref{tab:ranks}), led by Joplin (186.0 per million, 90\,\% interval 132.9--252.3) and Branson. None of the large hubs that dominate the raw counts appears. Read naively this says general-aviation fields are the most dangerous places in the system. Two features of the estimator argue against that reading. First, the exposure exponent is below one, so dividing by operations mechanically favours high-traffic airports; the ranking is therefore partly an artefact of the normalization it depends on. Second, the general-aviation share carries a rate ratio of 3.58 and these airports sit at the top of that distribution, so the covariate adjustment does not remove the traffic-mix difference, it prices it. What the table supports is the narrower claim that per movement, in the pattern-heavy general-aviation environment, incursions are far more frequent --- which is consistent with the FAA's own attribution of the majority of incursions to pilot deviations, and which says nothing about the probability that any one of them becomes a collision. That question belongs to the severity model. The FAA's own Runway Incursion Mitigation inventory \citep[July 2026 edition, 134 locations at 77 airports, of which 73 are designated hot spots;][]{faa2026riminventory} provides the natural comparison, and all 77 RIM airports fall inside the panel. Of them, 59 (76.6\,\%) have $\pi_i \geq 0.90$, so the count threshold and the shrinkage flag agree on three quarters of the RIM set. The disagreement runs in both directions and is informative. Nineteen RIM airports (24.7\,\%) have a 90\,\% rank interval that includes the median rank: once exposure and shrinkage are applied they cannot be distinguished from a typical airport, which is what selection on a raw count predicts. In the other direction 140 of the 199 flagged airports carry no RIM location at all, so the count rule misses more airports than it finds --- unsurprisingly, since a threshold of three incursions in a year is easiest to cross at airports that already have many movements. Empirical-Bayes and fully Bayesian shrinkage produce nearly the same ordering: the Spearman correlation between the two rate scales across the 565 airports is 0.999, and when the two rules are matched in size --- the 199 airports with the highest empirical-Bayes rates against the 199 with $\pi_i \geq 0.90$ --- the sets share 190 airports, with nine on each side of the disagreement. Unequal criteria would overstate the difference: an above-the-median empirical-Bayes rule flags 283 airports, so set against the 199 posterior flags it manufactures dozens of one-sided disagreements by arithmetic alone. What the full posterior adds is therefore not a different ranking but the uncertainty statement attached to it: the exceedance probability and the rank interval that say which flagged airports are separable from the median and which are not.

\begin{table}[htbp]
\centering
\caption{Airports with the highest posterior exceedance probability (top 15), with posterior rates, ranks and RIM status. RIM flag: Yes marks airports with at least one location in the July 2026 RIM inventory. The table is truncated for presentation; the full list of 565 airports (rates, exceedance probabilities and rank intervals) is released as supplementary data in \texttt{airport\_rates.csv}, which does not carry the RIM column --- the RIM flag shown here derives from the FAA inventory as stated.}
\label{tab:ranks}
\small
\begin{tabular}{@{}llrrrrrc@{}}
\toprule
Airport & Hub & Ops (M) & Incursions & Rate/$10^6$ (90\,\% PI) & $\pi_i$ & Rank (90\,\% PI) & RIM flag \\
\midrule
JLN & Non-hub & 0.39 & 79 & 186.0 (132.9--252.3) & $>0.999$ & 564 (562--565) & Yes \\
BBG & Non-hub & 0.11 & 21 & 171.4 (104.9--257.0) & $>0.999$ & 563 (555--565) & -- \\
MIC & Non-hub & 0.71 & 104 & 142.6 (104.6--190.0) & $>0.999$ & 561 (555--565) & -- \\
ASE & Non-hub & 0.72 & 84 & 123.4 (89.7--165.0) & $>0.999$ & 558 (547--564) & Yes \\
IDA & Non-hub & 0.56 & 70 & 120.7 (85.0--164.7) & $>0.999$ & 557 (544--564) & -- \\
NEW & Non-hub & 1.05 & 124 & 116.0 (85.0--153.1) & $>0.999$ & 556 (544--563) & Yes \\
MLI & Non-hub & 0.63 & 75 & 115.3 (83.1--155.5) & $>0.999$ & 556 (542--563) & -- \\
FCM & Non-hub & 1.78 & 205 & 113.4 (84.3--149.8) & $>0.999$ & 555 (543--563) & Yes \\
LSE & Non-hub & 0.36 & 44 & 112.3 (76.8--155.9) & $>0.999$ & 554 (537--563) & -- \\
SUN & Non-hub & 0.45 & 50 & 109.1 (75.7--150.3) & $>0.999$ & 553 (535--563) & -- \\
ADS & Non-hub & 1.83 & 173 & 106.6 (79.7--140.0) & $>0.999$ & 553 (539--562) & -- \\
MRI & Non-hub & 2.45 & 254 & 105.3 (79.8--136.8) & $>0.999$ & 553 (539--562) & Yes \\
VGT & Non-hub & 2.74 & 271 & 98.1 (74.0--127.8) & $>0.999$ & 549 (533--560) & Yes \\
BTR & Non-hub & 1.11 & 107 & 91.8 (67.0--121.3) & $>0.999$ & 544 (524--559) & -- \\
ARR & Non-hub & 1.14 & 109 & 91.6 (67.1--121.3) & $>0.999$ & 544 (524--559) & Yes \\
\bottomrule
\end{tabular}
\end{table}

\begin{figure}[htbp]
\centering
\includegraphics[width=0.49\linewidth]{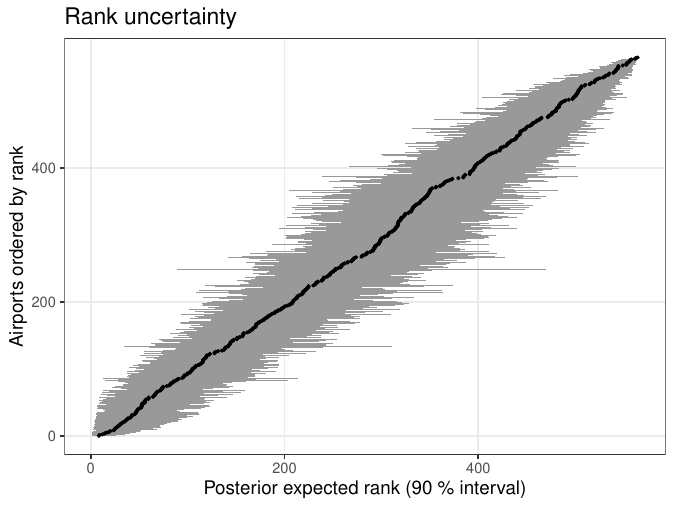}\hfill
\includegraphics[width=0.49\linewidth]{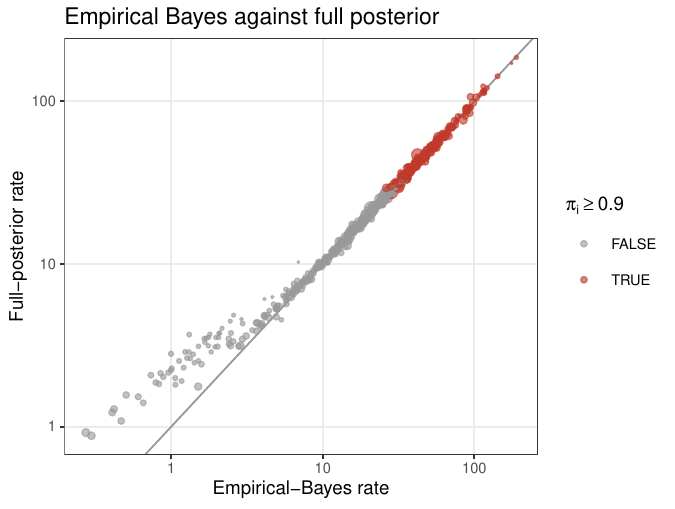}
\caption{Posterior rank distributions for FAA-towered airports (left) and agreement between empirical-Bayes and fully Bayesian rates (right). Rank intervals widen sharply below the top decile, indicating that finer ordering is not supported by the data; airports flagged by the raw-count RIM rule are distributed across the ranking.}
\label{fig:ranks}
\end{figure}

\subsection{RQ3: severity determinants and their relation to frequency}

The proportional-odds model is fitted to the 23{,}165 events that carry both an ICAO category (A--D) and a coded PD/OI/VPD type, at 510 airports in the panel (Table~\ref{tab:severity}). What separates severe from routine is overwhelmingly \emph{what kind of event it was}, not where it happened. An operational incident carries an odds ratio of 8.93 (8.23--9.72) for a more severe category relative to a pilot deviation, and a vehicle or pedestrian deviation 2.36 (2.04--2.74). Involvement of a part~121 air carrier raises the odds modestly, 1.29 (1.17--1.43); part~135 and part~129 operators are indistinguishable from part~91. Events whose first party is a vehicle or a pedestrian rather than an aircraft are markedly \emph{less} severe, 0.30 (0.26--0.35) and 0.17 (0.13--0.22), which is what the ICAO scale should produce: severity is graded by proximity to a collision between aircraft. Events with an unknown first-party operator --- 3{,}463 events, 15\,\% of the sample --- carry an extreme odds ratio of 0.020 (0.017--0.023) toward the less severe categories; because the unknown share tracks the reporting era rather than any property of the events (Section~\ref{sec:limitations}), this coefficient is a recording artefact, not an operator effect, and it is shown in Table~\ref{tab:severity} so that the covariate set is reported in full.

The airport random-effect standard deviation is $\sigma_v = 0.56$ (0.51--0.61), about half the dispersion of the frequency effect, so airports do differ in the severity mix they generate, but far less than in how often incursions occur at all.

The answer to RQ3 is the paper's most consequential result, and it is negative. Correlating the posterior-mean severity random effects of the 510 airports with their posterior-mean frequency rates gives a Spearman correlation of $-0.13$; propagating the uncertainty of both models --- computing the rank correlation draw by draw across paired posterior draws --- gives a posterior median of $-0.08$ with a 90\,\% interval of $-0.14$ to $-0.02$ and $\Pr(\rho < 0) = 0.99$. The correlation is thus small, probably negative, and bounded well away from anything that would make one ranking a usable proxy for the other. (The bivariate joint random-effects model named in the analysis plan was not run; these figures correlate two separately fitted models.) The airports where incursions happen most often per movement are not the airports where incursions are most likely to be severe --- if anything marginally the opposite, which is consistent with the composition result of Section~\ref{sec:results}: the high-rate airports are general-aviation fields, where the dominant event is a part~91 pilot deviation, the least severe combination in the model. It follows that an exposure-normalized frequency ranking, including the one this paper produces, carries little information about collision risk and should not be used as a ranking of it. A programme that allocates intervention by incursion rate is optimising a quantity that says almost nothing about where a catastrophic outcome is likely.

\begin{figure}[htbp]
\centering
\includegraphics[width=0.72\linewidth]{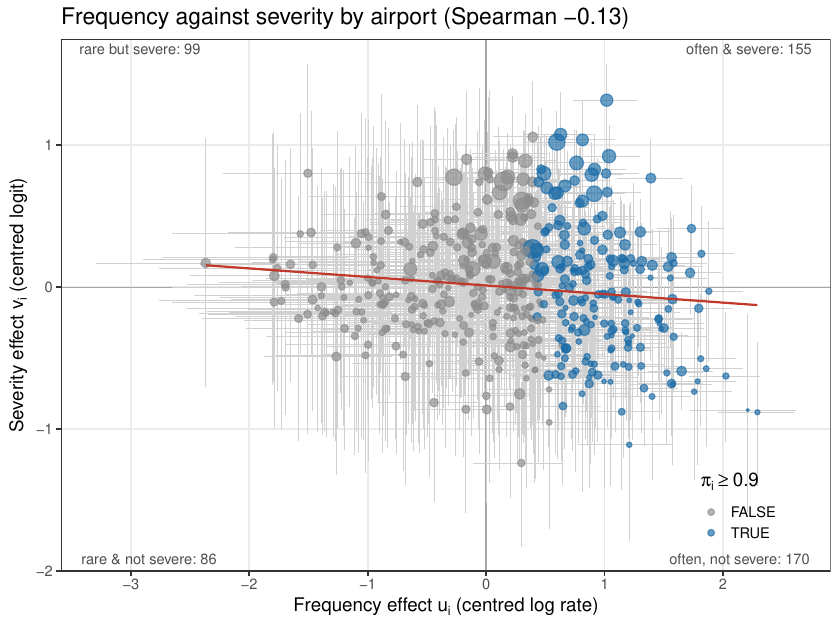}
\caption{Airport-level frequency and severity random effects. A weak or null correlation indicates that airports with elevated incursion frequency are not systematically the airports where incursions are more likely to be severe, consistent with the separation of determinants reported by Wilke et al.\ (2015).}
\label{fig:freqsev}
\end{figure}

Figure~\ref{fig:freqsev} makes the decoupling visible. Splitting the 510 airports at the mean of each effect leaves 155 that are both above-average in frequency and in severity, 170 frequent but not severe, 99 rare but severe and 86 low on both --- close to what independence would produce, and with the frequent-but-not-severe cell the largest. The airports flagged at $\pi_i \geq 0.90$ sit far to the right of the plot by construction, but they are spread across the whole vertical range: a high exposure-adjusted rate carries essentially no information about where that airport falls on severity.

The binary Category~A/B logistic model (Table~\ref{tab:severity}, column 2), admissible because the 233 Category~A and~B events exceed the threshold of 30 specified in the analysis plan, refines rather than overturns the ordinal picture. Operational incidents remain the dominant severity covariate in the extreme tail (odds ratio 4.77, 90\,\% interval 3.69--6.11), but two contrasts flip relative to the ordinal model: vehicle/pedestrian deviations are no longer elevated (0.99, 0.54--1.83), and part~121 involvement, mildly severity-increasing across the full A--D scale, is associated with \emph{fewer} A/B outcomes (0.63, 0.45--0.86) --- consistent with air-carrier events being more often classified C than D but rarely escalating to a near-collision. The airport random effect shrinks toward zero in the A/B tail ($\sigma_v = 0.40$, 90\,\% interval 0.11--0.61), so the data contain little evidence that specific airports systematically generate the most severe outcomes. One severity specification named in the analysis plan remains unrun: the partial-proportional-odds check.

\begin{table}[htbp]
\centering
\caption{Event-level severity models with 90\,\% intervals. Column (1): cumulative-logit (proportional-odds) model, odds ratios for a more severe ICAO category; severity is coded ascending, so an odds ratio above one shifts events toward more severe outcomes. Column (2): binary logistic model for a Category A/B versus C/D outcome (233 A/B events). The partial-proportional-odds check named in the analysis plan was not run.}
\label{tab:severity}
\small
\begin{tabular}{@{}lcc@{}}
\toprule
& (1) Ordinal A--D & (2) A/B logistic \\
\midrule
Operational incident (vs PD) & 8.93 (8.23--9.72) & 4.77 (3.69--6.11) \\
Vehicle/pedestrian deviation (vs PD) & 2.36 (2.04--2.74) & 0.99 (0.54--1.83) \\
Part 121 air carrier (vs part 91) & 1.29 (1.17--1.43) & 0.63 (0.45--0.86) \\
Part 135 operator (vs part 91) & 1.13 (0.99--1.28) & 1.39 (0.88--2.06) \\
Part 129 foreign operator (vs part 91) & 0.90 (0.73--1.13) & 0.94 (0.40--1.97) \\
Vehicle (vs part 91 aircraft) & 0.30 (0.26--0.35) & 0.78 (0.43--1.35) \\
Pedestrian (vs part 91 aircraft) & 0.17 (0.13--0.22) & 1.48 (0.59--3.59) \\
Operator unknown (vs part 91) & 0.020 (0.017--0.023) & 0.15 (0.07--0.29) \\
\multicolumn{3}{@{}l@{}}{\footnotesize Phase of flight, geometry and surveillance are not in the source export; see Section~\ref{sec:limitations}.} \\
\midrule
$\sigma_v$ (airport) & 0.56 (0.51--0.61) & 0.40 (0.11--0.61) \\
$\sigma$ (fiscal year) & 0.74 (0.57--1.02) & 0.42 (0.26--0.65) \\
Spearman $\rho(v_i,\ \log \lambda_i)$, posterior median & $-0.08$ ($-0.14$ to $-0.02$) & -- \\
Events & 23{,}165 & 23{,}165 (233 A/B) \\
\bottomrule
\end{tabular}
\end{table}

\subsection{RQ4: airport geometry}

RQ4 cannot be answered from the data assembled here and we state that rather than leave the section as an unfilled promise. The geometry block requires two FAA products that are not in the corpus: runway-crossing counts from the Airport Data and Information Portal, and the Runway Incursion Mitigation hot-spot inventory, which is published as PDF tables per report year. Neither is in the exposure panel, so the covariates named in the plan --- number of runway crossings and lagged hot-spot count --- were never estimated, and no coefficient for them is reported anywhere in this paper. Acquiring them is the single highest-value addition to this analysis, because geometry is the one candidate explanation for the residual between-airport dispersion of $\sigma_u = 1.05$ that survives exposure and traffic-mix adjustment.

\begin{figure}[htbp]
\centering
\fbox{\parbox{0.9\linewidth}{\centering\vspace{1.6cm}\textit{Not produced: the geometry block requires the FAA Airport Data and Information Portal runway-crossing counts and the Runway Incursion Mitigation hot-spot inventory, neither of which is in the corpus. The regression it would illustrate (RQ4) is likewise unestimated; see Section~\ref{sec:limitations}.}\vspace{1.6cm}}}
\caption{Partial effects of runway crossings and lagged hot-spot count on the exposure-adjusted incursion rate. The hot-spot effect is interpretable only as an association given that designation responds to past events.}
\label{fig:geometry}
\end{figure}

\subsection{Robustness}
\label{sec:robust-results}

Table~\ref{tab:robust} reports the stability of the exceedance set under the variations specified in advance, and the pattern is consistent: the statistical machinery is stable, the exposure definition is not.

The estimated quantities barely move. Across the refitted variants the between-airport standard deviation stays within 1.03--1.06, against 1.05 in the primary fit, and the crude national rate within 29.4--29.5 per million operations. Doubling or halving the prior scale on $\sigma_u$ changes the flagged set by two airports and one respectively, so the shrinkage is driven by the data rather than by the prior. Rebuilding the panel with the inclusion threshold lowered to 1{,}000 operations adds only six airport--years, all with zero incursions --- TAF-reporting towered airports rarely record a year below 5{,}000 operations --- and changes two flags; raising the threshold to 20{,}000, which drops the smallest airports from the panel entirely, changes 13. Excluding the pandemic years FY2020--FY2021 changes 17, and excluding the most recent fiscal year 9 --- reassuring, because it means the ranking is not an artefact of either the traffic collapse or of a fiscal year whose severity classification may still be incomplete.

The variation prompted by the type-code audit of the descriptive section behaves the same way. Restricting the outcome to the coded PD/OI/VPD types removes 1{,}062 events --- 311 and 333 of them in FY2016--FY2017 alone --- and lowers the crude rate to 28.3 per million. The airport-level machinery is indifferent to the exclusion: $\sigma_u$ moves from 1.05 to 1.06, the free exposure exponents are essentially unchanged (0.68 for non-hub and 0.81 for large-hub airports, against 0.68 and 0.80), the flagged set keeps 199 members, with four airports swapped in and four out, and the fiscal-year dispersion tightens from $\sigma_\gamma = 0.27$ to 0.24, which is exactly where the FY2016--FY2017 coding surge lived. The uncoded types inflate the national time series, not the cross-sectional comparison this paper is about.

The exception is the one that matters. Replacing total operations with itinerant-only operations as the exposure moves \textbf{66 airports} in or out of the flagged set, roughly a third of it, while leaving $\sigma_u$ and the national rate untouched. Together with the free-exponent result of Section~\ref{sec:results} --- $\rho$ between 0.68 and 0.83 across hub classes rather than 1 --- this locates the real fragility of any airport-level incursion ranking. It is not the count model, the priors or the shrinkage; it is the decision about what counts as exposure. An airport dominated by pattern work is a different denominator problem from a hub dominated by itinerant movements, and the two defensible choices disagree about a third of the airports one would flag. We therefore report the flagged set as conditional on the exposure definition and publish both, rather than presenting a single list.

The distribution-family comparison specified in the plan resolves in favour of the NB2. The zero-inflated negative binomial is statistically indistinguishable from it ($\Delta$ELPD $-2.0$, SE 3.6, below the plan's four-unit criterion) and estimates a zero-inflation probability of 0.7\,\% (90\,\% interval 0.2--1.4\,\%): the 40.8\,\% of zero airport--years need no separate zero process once overdispersion is modeled. The hurdle negative binomial is decisively worse ($\Delta$ELPD $-2237.6$, SE 55.0), as expected when structural and sampling zeros are forced apart in a panel whose zeros are ordinary small-rate outcomes. Extending the panel back to FY2002 (12{,}938 airport--years at the same 565 airports) leaves the between-airport dispersion essentially unchanged, $\sigma_u = 1.03$ (90\,\% interval 0.97--1.10), and estimates the FY2008 definitional step as a rate ratio of 2.2 (90\,\% interval 1.8--2.6); the crude rate over the extended window is 24.7 per million, reflecting the sparser pre-FY2008 recording that led the FAA to keep the two series apart. The lower bound is FY2002, not the FY2000 named in the analysis plan, because the frozen extract's earliest event is dated October 2001: FY2000 and FY2001 have no incursion records at all, and extending the exposure panel into them would add 1{,}117 airport--years of zeros that reflect absent data rather than absent incursions, inflating the apparent FY2008 step. (An earlier fit that included those two empty years estimated the step at 4.8 with a 1.7--12.8 interval; restricting the panel to the years the extract actually covers brings it down to 2.2 and tightens the interval, which is the honest version of this check.) Four elements of the plan's robustness program remain unrun and are reported as such rather than approximated: the latent-exposure treatment of non-towered airports, the leave-one-airport-out influence analysis, the GA-specific exposure alternative for pilot deviations, and the prior-scale variation on $\phi$ (the $\sigma_u$ prior variations are in Table~\ref{tab:robust}).

\begin{table}[htbp]
\centering
\caption{Sensitivity of key quantities to the variations specified in advance in the analysis plan, plus the coded-types variation added after the type-code audit of the descriptive section. $\Delta$flags is the number of airports entering or leaving the $\pi_i \geq 0.90$ set relative to the 199-airport flagged set of the full-length primary fit (Table~\ref{tab:frequency}); the variant models themselves are refitted with shorter chains (Appendix, Step~8). The $E \geq 1{,}000$ row rebuilds the panel from the TAF release with the lower threshold (9{,}588 airport--years; the six added airport--years all record zero incursions). Cells marked -- were not computed for that variant.}
\label{tab:robust}
\small
\begin{tabular}{@{}lcccc@{}}
\toprule
Variation & National rate & $\sigma_u$ & $\Delta$ELPD (SE) & $\Delta$flags \\
\midrule
Primary (NB2, airport ops offset) & 29.4 & 1.05 (0.98--1.11) & -- & -- \\
Poisson & 29.4 & -- & $-1509.5$ (250.5) & -- \\
Zero-inflated NB & 29.4 & 1.05 (0.99--1.11) & $-2.0$ (3.6) & -- \\
Hurdle NB & 29.4 & 0.62 (0.57--0.66) & $-2237.6$ (55.0) & -- \\
Itinerant-only exposure & 29.4 & 1.04 (0.98--1.11) & -- & 66 \\
Free exposure exponent (Eq.~\ref{eq:elastic}) & -- & -- & -- & Table~\ref{tab:frequency} \\
Excluding FY2020--21 & 29.5 & 1.03 (0.97--1.10) & -- & 17 \\
Excluding latest FY (2024) & 29.5 & 1.05 (0.98--1.12) & -- & 9 \\
Threshold $E \geq 1{,}000$ (panel rebuilt) & 29.4 & 1.05 (0.98--1.11) & -- & 2 \\
Threshold $E \geq 20{,}000$ & 29.4 & 1.03 (0.97--1.11) & -- & 13 \\
Coded types only (PD/OI/VPD) & 28.3 & 1.06 (1.00--1.13) & -- & 8 \\
Prior scale on $\sigma_u$ $\times 2$ & 29.4 & 1.05 (0.98--1.11) & -- & 2 \\
Prior scale on $\sigma_u$ $\times 0.5$ & 29.4 & 1.04 (0.98--1.11) & -- & 1 \\
Panel from FY2002 with FY2008 step & 24.7 & 1.03 (0.97--1.10) & -- & -- \\
\bottomrule
\end{tabular}
\end{table}

\section{Discussion}
\label{sec:discussion}

\subsection{Interpretation}

The primary output of this study is a set of posterior distributions from which any list of the most dangerous airports would have to be drawn, together with the evidence that such a list is or is not stable. If the estimated $\sigma_u$ were small relative to the sampling variation implied by $\phi$ and the typical count, the conclusion would be that exposure and traffic mix explain most of what distinguishes airports and that residual differences are weakly identified; if $\sigma_u$ is large, there is genuine airport-level heterogeneity that a national rate conceals. The estimates place this panel firmly in the second regime, $\sigma_u = 1.05$, which is the regime the RIM threshold and other count-based rules implicitly assume. Those rules, however, read individual counts as if the sampling noise around them were negligible, and the rank intervals show that it is not. The comparison between EB and fully Bayesian shrinkage is expected to matter mainly at low-volume towers, where the EB weight approaches one and the full posterior additionally reflects uncertainty in $\phi$ \citep{skaug2026beyond,efron1975data}.

\subsection{Alternative explanations}

Several mechanisms other than a difference in underlying risk could produce a high adjusted rate. Detection and reporting intensity differ with surveillance coverage, with the presence of a Runway Safety Action Team that is active in classification, and with the local interpretation of the C/D boundary; none of the three is observed in the estimated model (the surveillance indicators named in the analysis plan were not obtained), so all three mechanisms would load on the airport random effects, and detection bias remains inside $u_i$. Traffic mix is only partly captured by the GA share; the presence of a large flight school produces a concentration of student pilots that the ATADS user classes do not identify. Exposure error is airport-specific where local operations are estimated rather than counted. Finally, the ranking itself is conditional on the covariate set: an airport that ranks highly in the unadjusted model and near the median once GA share is included is not thereby exonerated, since the incursions still occur, but the appropriate mitigation is different. The released per-airport table therefore reports the crude rate alongside the covariate-adjusted empirical-Bayes and fully Bayesian rates, so that an airport's unadjusted position remains visible next to its adjusted one; a covariate-free hierarchical model was not estimated in this version.

\subsection{Implications}

For the FAA and airport operators the direct implication is that exceedance probabilities and rank intervals can replace or supplement raw-count thresholds in prioritizing RIM locations, and that before--after evaluations of geometry fixes should use the EB or hierarchical expected count as the baseline, as has been standard in road safety for two decades \citep{hauer2002estimating,persaud2007empirical}. The finding that severity and frequency determinants differ, if confirmed, supports the argument of \citet{stroeve2016strengthening} and \citet{wilke2015modelling} that outcome-based frequency monitoring should be complemented by risk-based assessment at airports where the geometry or the traffic mix makes a severe outcome more likely conditional on an incursion occurring. For data governance the study adds to the OIG's finding that integrated analysis is lacking \citep{oig2025runway}: the linkage of the incursion file to operations and geometry data is straightforward in principle and is here done reproducibly from public sources.

\subsection{Comparison with prior literature}

Our covariate effects can be compared directly with \citet{mathew2017statistical} and \citet{johnson2016statistical}, who find layout variables and traffic to be the main correlates of counts, and with \citet{omosebi2023investigating} on the dominance of pilot deviations; the differences are that exposure enters as an offset and that airport heterogeneity is modeled rather than absorbed into residuals. The severity results are comparable to \citet{biernbaum2012runway}, whose Volpe analysis on 2001--2010 data found controller-caused events roughly three times more likely to be severe and take-off-phase events 2.5 times more likely than taxi events; whether these ratios have persisted after the FY2008 definitional change and the expansion of surface surveillance is an empirical question this panel can answer. The ranking methodology follows \citet{goldstein1996league} without modification.

\subsection{Limitations}
\label{sec:limitations}

The panel covers only towered airports, so incursions at the several thousand non-towered public-use airports, where exposure is unobserved and reporting relies on pilot self-report, are excluded from the primary analysis; the latent-exposure sensitivity analysis that would impute their exposure is specified in the analysis plan but was not run (Section~\ref{sec:robust-results}). Reporting bias is intrinsic to the Category C/D series and cannot be separated from risk by fiscal-year effects alone; the surveillance covariate is a partial and endogenous proxy. Exposure is measured as operations, whereas the relevant exposure for a runway-crossing incursion is the number of crossings, which is not recorded. Severity classification of the most recent fiscal year may be incomplete at extraction. Geometry covariates depend on the coverage of ADIP and the RIM inventory, and hot-spot designation is endogenous even when lagged. The operator covariate in the severity model carries a recording artefact of its own: the share of events whose first-party operator is coded unknown moves with the reporting era, from 16\,\% in FY2008--FY2011 to 41--57\,\% in FY2012--FY2015 and then to about 2\,\% from FY2016 onward. The unknown category therefore proxies for the recording epoch as much as for any property of the event, its very low severity odds ratio (Table~\ref{tab:severity}) should be read as a data artefact rather than a finding about operators, and the remaining operator contrasts are estimated conditional on this compositional drift. Finally, the database export used for the incursion set is not itself publicly served through the API; reproducibility depends on the released export and the build scripts described in the appendix.

\subsection{Future work}

Extensions include a location-level model within airports using the RIM hot-spot geometry, a monthly rather than annual panel to test seasonality and its interaction with training activity, the addition of ASRS narratives as a self-reported comparison series, a cross-check of the incursion set against the wider investigated-occurrence corpus held at \site{} to identify the incursions that escalated to accidents or serious incidents, and a spatial model that borrows strength between airports in the same FAA region, where classification practice is shared. A European comparison awaits airport-level microdata that are not currently public \citep{eurocontrol2024gappri}.

\section{Conclusion}
\label{sec:conclusion}

This paper specifies and estimates the first multi-year, airport-level model of runway-incursion rates for FAA-towered airports with at least 5{,}000 annual operations, in which tower operations enter as an explicit exposure offset, airport heterogeneity is modeled hierarchically, and rankings are reported with posterior uncertainty. The design imports the empirical-Bayes and hierarchical-shrinkage methods that have long been standard for sparse crash counts in road safety and applies them to a dataset that has so far been analyzed mainly with raw counts, single cross-sections or fixed thresholds. The results (199 airports with exceedance probability above 0.9, a between-airport dispersion of $\sigma_u = 1.05$, and a frequency--severity rank correlation with posterior median $-0.08$) show that airport-level differences in incursion frequency are large and identifiable after exposure and traffic-mix adjustment, and that frequency carries little information about severity; both findings bear directly on how mitigation resources are allocated. The code, the frozen extract checksums and the build protocol are released so that the panel and every table can be rebuilt from FAA primary sources without access to the \ff{} database.

\section*{Data and code availability}

An interactive explorer of the underlying incursion extract --- raw event counts by airport and calendar year, without the exposure normalization, shrinkage estimates, exceedance probabilities or rank intervals of this paper --- is available at \url{https://himaxym.com/safety/runway-incursions}; the model outputs themselves are released in \texttt{airport\_rates.csv} in the archive described below. Everything needed to reproduce the paper is archived on Zenodo under the concept DOI \href{https://doi.org/10.5281/zenodo.21611388}{10.5281/zenodo.21611388}, which resolves to the latest version of the deposit: the event-level export \dataset{ri\_faa\_rws} (FAA Runway Safety data, US Government work, public domain), the estimation panels (\texttt{panel.csv}, \texttt{panel\_2000.csv}), the severity dataset (\texttt{severity.csv}), the per-airport table of shrunken rates, exceedance probabilities and rank intervals (\texttt{airport\_rates.csv}), the aggregate summaries (\texttt{incursions\_by\_year.csv}, \texttt{incursions\_by\_airport.csv}, \texttt{incursions\_airport\_year.csv}), the parsed RIM location inventory (\texttt{rim\_locations.json}, with the \texttt{pdftotext} layout extract it is built from), and a code archive (\path{p07-analysis-code-2026-11.zip}) that holds the build and estimation code (\texttt{build\_*.mjs}, \texttt{fit\_*.R}, \texttt{resume\_model.R}, \texttt{compute\_spearman\_draws.R}, the figure scripts, the column crosswalk \texttt{schema/ri\_columns.yaml}, the \texttt{Makefile} and the analysis plan), the summaries of the published fits, and \texttt{check\_numbers.mjs}, which recomputes the numbers reported here and checks them against this manuscript; its \texttt{REPRODUCE.md} lists the steps. The export's record count, extraction date and SHA-256 checksum are recorded in the appendix. The runway-incursion set is not served by the public \site{} API; the appendix protocol rebuilds the panel from the FAA primary sources alone (the ASIAS RWS export and the TAF bulk download), so no access to the \site{} database is required. Tower operations are from the FAA Terminal Area Forecast bulk release, whose historical years are the ATADS/OPSNET counts (public domain). Airport metadata are retrieved from \apiep{GET} \url{https://himaxym.com/api/v1/data/airports/{ident}} under the licence terms returned by \apiep{GET /sources}; the OpenAPI description is at \url{https://himaxym.com/api/v1/data/openapi.json}. The RIM location inventory is from the FAA RIM program's inventory of airport locations, July 2026 edition \citep{faa2026riminventory} (public domain).

\section*{Author contributions and competing interests}

The sole author designed the study, built the data pipeline, wrote the analysis code and the manuscript. Competing interests: the author operates \site{} (\ff), the aggregation platform from which the incursion export and airport metadata are drawn. The platform is used here as a data source with the stated methodology; no revenue depends on the findings, and all inputs are traceable to FAA primary files that are released with the paper.

\appendix

\section{Reproducibility protocol}
\label{app:repro}

\paragraph{Step 1: incursion extract.} From the \ff{} database, export the FAA Runway Safety table in full with \texttt{SELECT * FROM ri\_faa\_rws} --- no date restriction: the FY2008--FY2024 analysis window is imposed at the build stage (Step~7), and the FY2002--FY2007 records feed the extended panel of Section~\ref{sec:robust-results} --- to \texttt{ri\_faa\_rws.csv}; record the record count (35{,}308 data records; the file has 35{,}371 non-header physical lines because 62 records contain a line break inside a quoted field --- 63 embedded line breaks in total, one record carrying two --- so a quote-aware CSV parser is required), the SHA-256 checksum (\texttt{0cba71db\dots67247d5}) and the date the export was pulled (2026-07-14). The equivalent primary-source route is the ASIAS Runway Incursion query tool (all airports, all fiscal years, all severities), exported to CSV; column names are mapped by the crosswalk in \texttt{schema/ri\_columns.yaml}, released with the paper.

\paragraph{Step 2: operations extract.} Download the FAA Terminal Area Forecast bulk release (file \path{APO100_TAF_Final_2025.zip}) and read \texttt{AirportsOperations.xlsx} (operations by airport--fiscal-year in the six itinerant and local user-class fields of Table~\ref{tab:variables}) and \texttt{Airports.xlsx} (inventory with \texttt{ATCT\_FLAG} and \texttt{HUB\_SIZE}). Keep only rows with \texttt{scenario}~$=0$: each airport--year also appears under forecast-scenario codes, and retaining them multiply-counts exposure. Trim trailing whitespace from \texttt{LOCID} before joining; the TAF tables carry it.

\paragraph{Step 3: airport metadata (optional).} The panel build takes the towered-airport flag (\texttt{ATCT\_FLAG}) and hub class (\texttt{HUB\_SIZE}) from the TAF \texttt{Airports.xlsx} inventory of Step~2; no API access is needed to reproduce any table or figure in this paper. To retrieve coordinates or cross-check identifiers, one can additionally call, for each distinct airport identifier,
\begin{verbatim}
curl -H "Authorization: Bearer $KEY" \
  "https://himaxym.com/api/v1/data/airports/{ident}"
\end{verbatim}
and store \texttt{ident}, \texttt{icao}, \texttt{hub\_class}, \texttt{tower\_type}, \texttt{runways[]} and coordinates. Rate limits (60/min on the free tier) require roughly 10 minutes for the 565 identifiers in the panel; responses are cached.

\paragraph{Step 4: identifier crosswalk.} Incursion records use FAA location identifiers; ATADS uses facility identifiers, which differ where a tower serves more than one airport or where identifiers changed. Map through the API \texttt{ident}/\texttt{icao} pair, then by direct match on the location identifier after trimming trailing whitespace, which the Terminal Area Forecast tables carry. No manual override list was needed: 27{,}079 of the 27{,}242 in-window incursions, 99.4\,\%, matched a towered airport in the exposure frame. Unmatched incursion records are reported and excluded.

\paragraph{Step 5: deduplication.} Collapse records with identical \texttt{EVENT\_ID}; then collapse records with identical (\texttt{airport\_id}, \texttt{event\_date}, \texttt{event\_type}, \texttt{runway}) and severity differing by at most one category, keeping the more severe category and flagging the collapse. In this export no collapsing was performed and none was warranted: there are no repeated \texttt{EVENT\_ID} values in the analysis window. The tuple rule was not applied because the runway field (\texttt{EVENT\_TKOF\_LNDG\_DESC}) is uninformative --- blank, \texttt{UNK}, \texttt{NA}, \texttt{N/A} or the literal string \texttt{NULL} --- for 6{,}823 of the 27{,}242 in-window records, 25.0\,\% (4{,}933 blank, 450 \texttt{UNK}, 929 \texttt{NA}, 334 \texttt{N/A}, 177 \texttt{NULL}, counted with a quote-aware CSV parse), which makes the key degenerate for a quarter of the file; applying it regardless would merge 1{,}383 records into earlier ones across 887 duplicate-key groups, most of which are separate events at the same airport on the same day rather than duplicates. The counts collapsed at each step are therefore zero and zero.

\paragraph{Step 6: derived variables.} Fiscal year as above; $E_{it}$ = itinerant total + local total; GA share = (\texttt{itn\_ga} + \texttt{loc\_ga}) / $E_{it}$; local share = (\texttt{loc\_ga} + \texttt{loc\_mil}) / $E_{it}$; hub class from \texttt{HUB\_SIZE}. The surveillance, RWSL and geometry covariates named in the analysis plan were not constructed (Section~\ref{sec:limitations}).

\paragraph{Step 7: panel filters.} Retain airport--years with at least 5{,}000 operations. The exposure release is annual rather than monthly, so the ten-reporting-months rule in the analysis plan does not apply and the operations threshold carries the inclusion decision alone; retain FY2008 onward for the primary model and FY2002 onward --- the first fiscal year present in the extract --- for the extended panel.

\paragraph{Step 8: estimation.} R 4.6.1, \texttt{brms} 2.23.0 on the \texttt{rstan} 2.32.7 backend, \texttt{loo} 2.10.1; seed 20260827 (the posterior predictive checks draw without a fixed seed, so their replicated values vary slightly between reruns); \texttt{adapt\_delta} = 0.95 throughout, except 0.9 for the ordinal severity model. Chain settings by model group: the primary NB2 and the Poisson comparison use four chains of 2{,}000 sampling iterations after 1{,}000 warm-up (8{,}000 draws); the type-specific and free-exponent models eight chains of 1{,}000 after 1{,}000; the severity model eight chains of 750 after 750; the family-comparison, A/B-logistic and extended-panel fits six chains of 750 after 750; the robustness refits of Table~\ref{tab:robust} eight chains of 750 after 750; and the coded-types sensitivity fits six chains of 750 after 750. No fit produced divergent transitions. Low bulk-ESS warnings were emitted by the free-exponent model of Table~\ref{tab:frequency}, column (2) --- the model on which the rejection of the offset assumption rests --- by the severity model, by the zero-inflated, hurdle and extended-panel fits, and by the short-chain robustness refits; the corresponding interval summaries in Tables~\ref{tab:frequency} (column 2) and \ref{tab:severity}--\ref{tab:robust} should therefore be read as approximate. The EB benchmark uses \texttt{MASS::glm.nb} for the safety-performance function. The model-fitting scripts record the exact settings of each run.

\paragraph{Step 9: outputs.} Tables~\ref{tab:descriptives}--\ref{tab:robust} and Figures~\ref{fig:national}--\ref{fig:geometry} are generated by \texttt{make all} with the released \texttt{Makefile} (build scripts, model fits, figure scripts, then \texttt{latexmk}); the full per-airport table with rates, exceedance probabilities and rank intervals is written to \texttt{data/airport\_rates.csv}.

\bibliographystyle{unsrtnat}
\bibliography{refs}

\end{document}